\documentclass[
 reprint,
superscriptaddress,
 amsmath,amssymb,
prl,
floatfix,
]{revtex4-2}

\usepackage{graphicx}
\usepackage{dcolumn}
\usepackage{bm}
\usepackage{color}
\usepackage{float}
\usepackage[colorlinks,urlcolor=blue,citecolor=blue,linkcolor=blue]{hyperref}

\begin{document}

\preprint{topo}

 \title{Enhancing ground state population and macroscopic coherence of room-temperature WS$_2$ polaritons through engineered confinement}

\author{M.~Wurdack}
 \email{matthias.wurdack@anu.edu.au}
 \affiliation{ARC Centre of Excellence in Future Low-Energy Electronics Technologies and Department of Quantum Science and Technology, Research School of Physics, The Australian National University, Canberra, ACT 2601, Australia}

\author{E.~Estrecho}%
 \affiliation{ARC Centre of Excellence in Future Low-Energy Electronics Technologies and Department of Quantum Science and Technology, Research School of Physics, The Australian National University, Canberra, ACT 2601, Australia}
 
\author{S.~Todd}
 \affiliation{ARC Centre of Excellence in Future Low-Energy Electronics Technologies and Department of Quantum Science and Technology, Research School of Physics, The Australian National University, Canberra, ACT 2601, Australia}
 
\author{C.~Schneider}
\affiliation{Institut f\"ur Physik, Carl von Ossietzky Universit\"at Oldenburg, Ammerl\"ander Heerstra{\ss}e 114-118, 26126 Oldenburg, Germany}%

\author{A.~G.~Truscott}%
 \affiliation{Department of Quantum Science and Technology, Research School of Physics, The Australian National University, Canberra, ACT 2601, Australia}
 
 \author{E.~A.~Ostrovskaya}
  \email{elena.ostrovskaya@anu.edu.au}
 \affiliation{ARC Centre of Excellence in Future Low-Energy Electronics Technologies and Department of Quantum Science and Technology, Research School of Physics, The Australian National University, Canberra, ACT 2601, Australia}

\begin{abstract}
Exciton-polaritons (polaritons herein) in transition-metal dichalcogenide monolayers have attracted significant attention due to their potential for polariton-based optoelectronics. Many of the proposed applications rely on the ability to trap polaritons and to reach macroscopic occupation of their ground energy state. Here, we engineer a trap for room-temperature polaritons in an all-dielectric optical microcavity by locally increasing the interactions between the WS$_2$ excitons and cavity photons. The resulting confinement enhances the population and the first-order coherence of the polaritons in the ground state, with the latter effect related to dramatic suppression of disorder-induced inhomogeneous dephasing. We also demonstrate efficient population transfer into the trap when optically injecting free polaritons outside of its periphery.
\end{abstract}
\maketitle

\textit{Introduction.}-- Excitons (bound electron-hole pairs) in atomically-thin layers of transition metal dichalcogenide crystals (TMDCs) have a large binding energy \cite{Chernikov2014} and interact strongly with light \cite{Mak2010, Splendiani2010, TMDC2012}. Excitons strongly interacting and hybridising with photons in an optical microcavity give rise to polaritons \cite{Weissbuch1992,Microcavities,Kasprzak2006,Deng2010,Carusotto2013,Byrnes2014,Schneider2018}. Hence, creating a local minimum in the potential energy of either the photon or exciton component of polaritons can lead to their spatial confinement \cite{Kaitouni2006}. In particular, polariton trapping through modification of the photon energy in patterned microcavities \cite{Schneider2016}, has enabled many proof-of-principle demonstrations of potential applications, such as polariton lasers \cite{Bajoni2008,Schneider2012}, topological insulators \cite{Klembt2018}, sensors \cite{Sturm2014}, and non-classical state generators \cite{Imamoglu2019,Munoz2019}.

Recently, enhanced coherence was demonstrated for TMDC polaritons \cite{Zhao2020,Shan2021,Anton2021,Wurdack2021} and attributed to either strong motional narrowing \cite{Wurdack2021} or the onset of polariton condensation \cite{Zhao2020,Shan2021,Anton2021}. At room temperature \cite{Zhao2020,Shan2021,Wurdack2021}, these observations relied on accidental traps arising from strain, fractures in the monolayer, or air gaps in the cavity spacer, all of which modulate the energies of the excitons and/or cavity photons. Controlled engineering of a potential landscape for TMDC polaritons remains a challenge, since the monolayers are easily damaged by most fabrication techniques \cite{Yun2021}. So far, such engineering has only been achieved in an open cavity design \cite{Lackner2021}.

In this work, we create a trap for WS$_2$ polaritons by locally altering the strength of interactions between the excitons and the photons in a microcavity. This is achieved by stacking a small monolayer on top of a WS$_2$/Ga$_2$O$_3$ heterostructure \cite{Wurdack2020}, which increases the exciton-photon coupling strength \cite{Rupprecht2020} and creates an effective potential well. By studying the photoluminescence (PL) of the polaritons inside the well, in comparison to the free polaritons with similar excitonic fraction, we observe a significant increase of the ground state population in the trap. Power-dependent measurements link this increase to enhanced parametric scattering in the polariton relaxation bottleneck \cite{Tassone1997,Tassone1999,Savvidis2000,Muller2000,Tartakovskii}. Additionally, we find that spatial and temporal first-order coherence of polaritons in the ground state of the trap is significantly enhanced compared to the low-energy states of the free polaritons, and reaches the values comparable to those previously linked to the onset of bosonic condensation in TMDCs at room temperature \cite{Zhao2020,Shan2021}. However, our results strongly suggest that the enhanced coherence is due to coupling of the short-living excitons to the longer-lived cavity photons, which reduces both the homogeneous decoherence and inhomogeneous dephasing \cite{Wurdack2021}, with the trapping further suppressing these effects. Finally, our sample design allows us to efficiently populate the trap with partially coherent polaritons while exciting them outside the trap.

\begin{figure}[ht!]
\centering
\includegraphics[width=8.6cm]{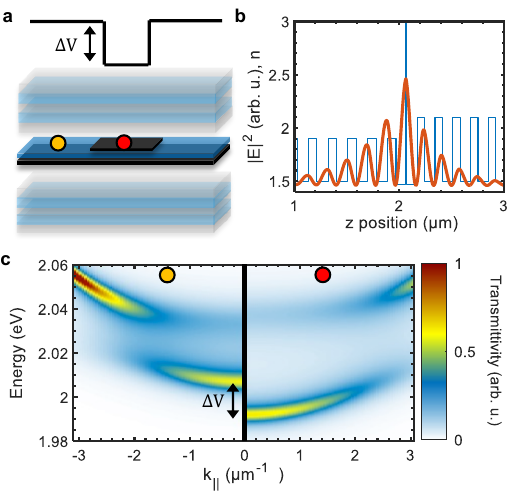}
\caption{(a) Sample design for creating a trap for WS$_2$ polaritons by placing a small monolayer (red dot) on top of a WS$_2$/Ga$_2$O$_3$ heterostructure (orange dot), sandwiched between DBR mirrors. (b) Calculated distribution of the light field inside the microcavity. The different materials are represented by their respective dielectric constants, $n$. (c) Calculated angle-resolved transmission spectra of the structure at the position of (left) the large monolayer and (right) the small monolayer, marked with the orange and red dots, respectively. The energy difference, $\Delta V$, corresponds to the effective trapping potential at the position of the small monolayer in (a).}
\label{fig:Fig1}
\end{figure}

\textit{Engineering polariton confinement}.-- Fast energy exchange between excitons and cavity photons with the respective energies $E_X$ and $E_C$ results in the emergence of the polariton dispersion branches \cite{Microcavities}: $E_{\rm LP/UP}(k_{||}) = (1/2)[E_{X}(k_{||}) + E_{C}(k_{||})\pm \sqrt{E_R^2 + \delta(k_{||})^2}]$. The exciton-photon detuning, $\delta=E_C-E_X$, and the Rabi splitting, $E_R$, define the excitonic and photonic fraction of the polariton through the respective Hopfield coefficients $|X|^2=(1/2)[1+\delta/\sqrt{E_R^2+\delta^2}]$, $|C|^2=1-|X|^2$. A faster energy exchange increases the splitting between the polariton branches and decreases the lower polariton energy at zero in-plane momentum $E_{\rm LP}(k_{||}=0)$. This can be achieved by placing $N$ exciton-hosting layers in an E-field maximum of the cavity with the Rabi splitting growing as \cite{Microcavities}: $E_R\propto \sqrt{N}$. Hence, when a small WS$_2$ monolayer, with the size comparable to the polariton de Broglie wavelength, is placed onto a larger WS$_2$ monolayer, the increased Rabi splitting at the double-layer position can induce an effective trapping potential comparable to $E_R$ for the lower polariton. However, interlayer interactions can affect the optical response of the monolayers \cite{Zhou2018}, create interlayer excitons \cite{Das2020}, and weaken the coupling of the intralayer excitons to the photons. This can be avoided by using a tunnel barrier, such as a hBN spacer \cite{Menon2019,Rupprecht2020} between the monolayers. Here, we use a new passivation material, ultrathin Ga$_2$O$_3$ glass, which enables reliable fabrication of large-area samples \cite{Wurdack2020}. 

The value of $E_{\rm LP}(k_{||}=0)$ defines the effective potential energy for lower polaritons. To estimate the potential energy difference between polaritons in a single layer and in a double layer (see Fig.~\ref{fig:Fig1}a) and to account for the local elongation of the microcavity, we perform transfer matrix calculations \cite{Microcavities} using the linewidth and the oscillator strength of the excitons extracted from the experiment \cite{SM}. The energy exchange between the photons and the excitons is maximized when the monolayers are placed in the E-field maximum (see Fig.~\ref{fig:Fig1}b). Simulated transmission spectra for this configuration show the anticrossing of the transmission maxima along $k_{||}$ for both single and double layers (see Fig.~\ref{fig:Fig1}c), signalling the strong exciton-photon coupling regime. The Rabi splitting is significantly larger in the double layer. At this detuning, the potential energy step from the single layer to the double layer, $\Delta V \approx 15~\mathrm{meV}$, is comparable to $E_R \approx 25~\mathrm{meV}$, and is sufficient to trap polaritons.

\begin{figure}[ht!]
\centering
\includegraphics[width=8.6cm]{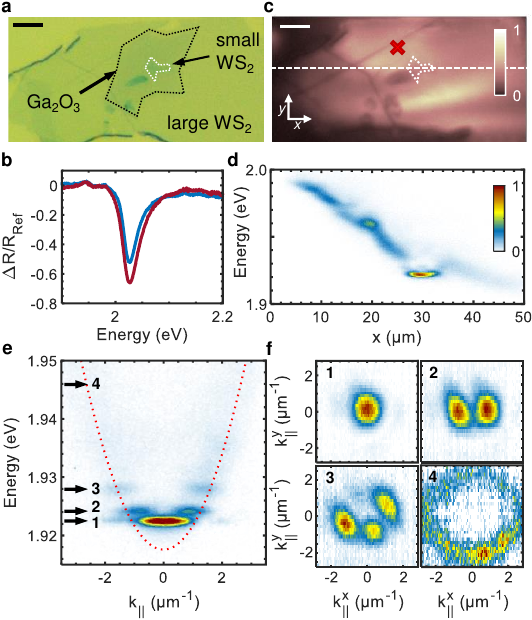}
\caption{(a) Microscope image of the WS$_2$/Ga$_2$O$_3$/WS$_2$ heterostructure on top of a DBR substrate. The large monolayer spans $\sim 20~\mathrm{\mu m}$ in $y$ direction and $>50~\mathrm{\mu m}$ in $x$ direction, and the top monolayer is $\sim 4~\mathrm{\mu m}$ in both directions. The scale bar is $5~\mathrm{\mu m}$. (b) Reflectivity spectra of the (blue) WS$_2$/Ga$_2$O$_3$ and (red) WS$_2$/Ga$_2$O$_3$/WS$_2$ heterostructure. (c) PL image of the complete microcavity excited with a large Gaussian laser spot (${\rm FWHM}\approx 30~\mathrm{\mu m}$, $\lambda = 532 ~\mathrm{nm}$). The scale bar is $5~\mathrm{\mu m}$.  (d) Spatially resolved PL spectrum at $k_{||}\approx 0$ along the dashed line in (c). (e) Angle-resolved PL spectrum at $x\approx30~\mathrm{\mu m}$. (f) Polariton mode tomography in the trap ($x\approx30~\mathrm{\mu m}$) at the energies of the (1) ground, (2) first excited, (3) second excited, and (4) continuum state $\sim15~\mathrm{meV}$ above the potential well, marked with the arrows in (e).}
\label{fig:Fig2}
\end{figure}

After fabricating the WS$_2$/Ga$_2$O$_3$/WS$_2$ heterostructure on a highly reflective distributed Bragg reflector (DBR) substrate ($R>99.99~\%$ at $\lambda = 600~\mathrm{nm}$) \cite{ANNF}, see Fig.~\ref{fig:Fig2}a, the microcavity is completed by mechanical transfer of the top DBR \cite{Rupprecht2021,Wurdack2021}. White-light reflectivity measurements performed on the heterostructure prior to the microcavity assembly show a decreased reflectivity dip at the position of the double layer (see Fig.~\ref{fig:Fig2}b) due to the increased exciton-photon interaction. The PL image in Fig.~\ref{fig:Fig2}c,  obtained by exciting the sample with a large continuous-wave (cw) laser spot, shows a strong monolayer emission, with dark areas arising from the bulk WS$_2$ transferred together with the small monolayer. Angle-resolved PL measurements at different positions confirm that the whole sample is operating in the strong coupling regime \cite{SM}. In the single-layer region, the PL reveals a free polariton dispersion (shown in \cite{SM}) with $E_R \approx 28~\mathrm{meV}$ and a range of detuning values corresponding to highly photonic polaritons, $|X|^2\ll 1$.

The spatially resolved PL spectrum at $k_{||}\approx 0$ along the dashed line in Fig.~\ref{fig:Fig2}c \cite{SM} allows us to trace the potential energy profile for the polaritons in $x$-direction (see Fig.~\ref{fig:Fig2}d) \cite{Pieczarka2019,Wurdack2021}. The energy gradient arises due to a tilt of the top DBR leading to a gradient of $E_C$. Between $x=27~\mathrm{\mu m}$ and $33~\mathrm{\mu m}$, at the position of the small monolayer, the enery at $k_{||}\approx 0$ drops, signalling the formation of a potential well with the depth $\Delta V \approx 10~\mathrm{meV}$, in agreement with the transfer matrix calculation \cite{SM}. The spectrum is discretized at this position (see Fig.~\ref{fig:Fig2}e). It becomes continuous above the potential well, and can be fitted to a free polariton dispersion with $E_R \approx 33~\mathrm{meV}$, and $\delta = -107 ~\mathrm{meV}$, corresponding to $|X|^2=0.022$ \cite{comparison}. PL images at the discrete energies in the trap (see Fig.~\ref{fig:Fig2}f) reveal 2D-confined modes. The intensity profiles (polariton densities) are skewed due to the potential gradient, with the same asymmetry observed in the continuum (Fig.~\ref{fig:Fig2}f). These observations agree with modelling of a tilted box trap shaped as the small monolayer \cite{SM}. 

\begin{figure}[ht!]
\centering
\includegraphics[width=8.6cm]{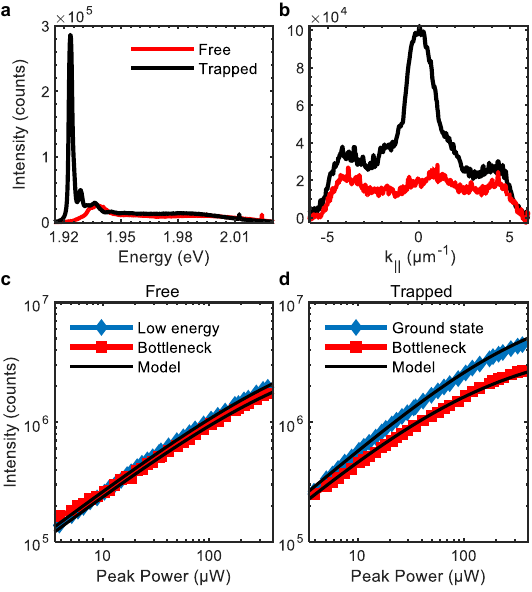}
\caption{(a) Energy and (b) momentum-resolved PL of the (red) free and (black) trapped polaritons. (c,d) Power-dependent intensities at low energy and bottleneck regions for the (c) free and (d) trapped polaritons extracted from the angle-resolved PL spectra \cite{SM}. Black lines are the polariton intensities calculated using the model in \cite{SM}.}
\label{fig:Fig3}
\end{figure}

{\em Enhanced ground state population.}-- The PL spectra integrated along energy or momentum (see Fig.~\ref{fig:Fig3}a,b) show an order of magnitude enhancement of the polariton population in the ground state of the trap compared to the low-energy states of free polaritons at the position marked in Fig.~\ref{fig:Fig2}c. In addition, the states at $k_{||}\approx \pm4.4~\mathrm{\mu m^{-1}}$ are strongly populated. This region around the inflection point of the dispersion is part of the energy relaxation bottleneck \cite{Tassone1997,Tassone1999,Muller2000,Tartakovskii,Kasprzak2008,Deng2010,Byrnes2014}. While the bottleneck plays an important role in the polariton energy relaxation in quantum-well based microcavities, its role in the TMDC-based systems remains unexplored. In our sample, the pronounced bottleneck population implies suppressed energy relaxation and relatively long polariton lifetimes, enabling them to accumulate before decaying radiatively. Although elevated temperatures suppress the bottleneck in quantum-well systems \cite{Tassone1997,Tartakovskii,Savvidis2002}, our observation of the bottleneck at room temperature points to weak interactions of the WS$_2$ polaritons with LO-phonons at large negative $\delta$ \cite{Lengers2021}. 

To gain an insight into the polariton energy relaxation, we compare the populations at the bottleneck and in the low-energy region with increasing pump power in a quasi-cw, pulsed regime \cite{SM}. Figures \ref{fig:Fig3}c,d show the power-dependent intensities proportional to the populations of the free and trapped polaritons in their respective bottleneck and low-energy states at a fixed duration of the pump pulse ($50~\mathrm{\mu s}$). The intensity of both regions saturates at higher powers, which was previously attributed to Auger recombination of excitons in monolayer WS$_2$ \cite{Mouri2014,Hoshi2017,Zipfel2020}. The populations of free polaritons in both regions increase with pump power with an equal rate (see Fig. \ref{fig:Fig3}c), while the growth rate of the ground state population in the trap is significantly enhanced (see Fig. \ref{fig:Fig3}d). This behaviour can be modelled with a 3-level system, which includes a high-energy excitonic reservoir injected by the pump, the bottleneck region, and the ground state \cite{SM}. The relative populations in these states are governed by the phonon-driven energy relaxation with the rates dictated by the density of states \cite{Muller2000}, spontaneous parametric polarion-polariton scattering from the bottleneck into the ground state \cite{Tartakovskii,Savvidis2000,Savvidis2002,Deng2002,Wang2015}, and exciton annihilation (e.g., Auger recombination) in the excitonic reservoir \cite{SM}. The observed enhancement of the growth rate of the ground state population is reproduced when the polariton-polariton scattering rate for the trapped polaritons exceeds that for the free polaritons by one order of magnitude (see Fig. \ref{fig:Fig3}c,d). This points to an increased probability of the polariton-polariton scattering in the trap, leading to a large population in the ground state (Fig.~\ref{fig:Fig3}a,b) and enhanced ground state emission \cite{Tartakovskii}. However, further studies are needed to confirm this. 

\begin{figure}[ht!]
\centering
\includegraphics[width=8.6cm]{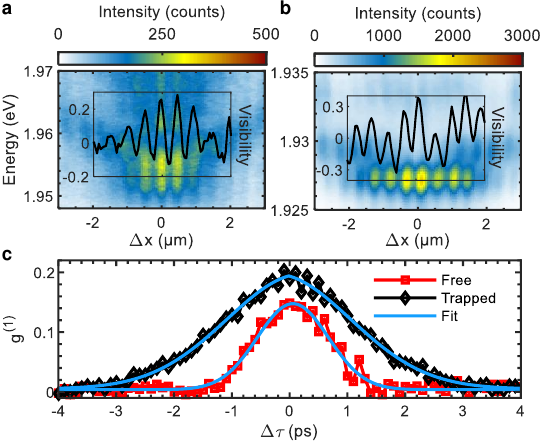}
\caption{(a,b) Spectrally-resolved Michelson interference maps of the (a) free and (b) trapped polaritons \cite{SM}. The overlays are the normalized interference fringes at the energies marked by the $x$-axis position, with the amplitude corresponding to their visibility and $g^{(1)}(\Delta x)$ \cite{SM}. (c) Temporal first-order coherence $g^{(1)}(\Delta \tau)$. Solid lines are the fits accounting for inhomogeneous and homogeneous decoherence (see text).}
\label{fig:Fig4}
\end{figure}

{\em Enhanced macroscopic coherence.}-- Spectrally and spatially resolved first-order coherence $g^{(1)}(\Delta x,\Delta\tau)$ of the polaritons is measured by using a modified Michelson interferometer \cite{Shan2021,Wurdack2021} \cite{SM}. When overlapping the spatially resolved PL spectrum with its flipped image at $\Delta \tau=0$, we observe clear interference fringes (see Figs.~\ref{fig:Fig4}a,b). The normalized fringes \cite{SM}, with the magnitude proportional to $g^{(1)}(\Delta x)$, extend over a larger region for the trapped polaritons compared to the free polaritons in their respective low-energy states. This is expected since the spatial extent of a trapped polariton wavefunction \cite{SM} is larger than its thermal de Broglie wavelength $\lambda_{th}\approx1.36~\mathrm{\mu m}$, calculated with the effective mass extracted from the fits in Fig.~\ref{fig:Fig2}f and in \cite{SM}. However, the magnitude of partial coherence at $\Delta x\approx0$ is similar for the trapped and the free polaritons and remains constant over a large range of pump powers \cite{SM}. The absence of correlation with the enhancement of the ground state population (Fig.~\ref{fig:Fig3}f) suggests that the enhanced partial coherence of the WS$_2$ polaritons is inherited from their photonic component. 

The temporal coherence of the polaritons is measured by applying a time delay $\Delta\tau$ to the flipped image \cite{SM}, and extracting the fringe visibility around $\Delta x=0$, see Fig. \ref{fig:Fig4}c. To quantify the decay of coherence, we fit the data to the coherence function accounting for inhomogeneous, $T_{I}$, and homogeneous, $T_{H}$, decoherence times \cite{Reimer2016}: $g^{(1)} (\Delta \tau) = g^{(1)} (0)\exp\left[-(\pi/2)\left(\Delta\tau/T_{I}\right)^2 - \left|\Delta\tau\right|/{T_{H}}\right]$. The fitting shows that the free polaritons are mainly affected by inhomogeneous decoherence, with $T_{I} = (1084\pm149)~\mathrm{fs}$ and negligible $T_{H}$. Indeed, the low-energy states of the free polariton continuum are affected by fluctuations of the cavity photon energy, i.e., due to the potential gradient (see Fig.~\ref{fig:Fig3}d), which causes linewidth broadening and inhomogeneous dephasing. In contrast, for the trapped polaritons in their ground state, $T_{I} = (2334\pm 166)~\mathrm{fs}$ and $T_{H} = (6960\pm3747)~\mathrm{fs}$. The large $T_{I}$ points to the suppression of inhomogeneous dephasing due to exciton energy fluctuations \cite{Wurdack2021} and to insensitivity of the discrete state to photon energy fluctuations. The narrow linewidth of the trapped polariton states, Fig.~\ref{fig:Fig3}a, is therefore mainly governed by their lifetime, $\tau_{p}$, which can be extracted from $T_{H}$. The resulting $\tau_{p}$ of several picoseconds exceeds the radiative lifetime of the WS$_2$ excitons by at least an order of magnitude \cite{Poellmann2015,Selig2016}, meaning that it is the long lifetime of the cavity photon, $\tau_C$, that determines the lifetime and hence coherence time of highly photonic polaritons in a high-quality microcavity: $\tau_{p}^{-1}\propto |C|^2 \tau_{C}^{-1}$. The coherence times for both free and trapped polaritons are similar to those reported previously \cite{Zhao2020,Shan2021}, and are comparable to $\tau_{\rm p}$. One needs to exercise caution when interpreting these coherence times as a signature of bosonic condensation \cite{Zhao2020,Shan2021} because in conventional systems (e.g., quantum wells and organic semiconductors) the condensate coherence time is much larger than $\tau_{p}$ under quasi-cw excitation \cite{Askitopoulos2019,Betzold2020,Orfanakis2021}.

\begin{figure}[ht!]
\centering
\includegraphics[width=8.6cm]{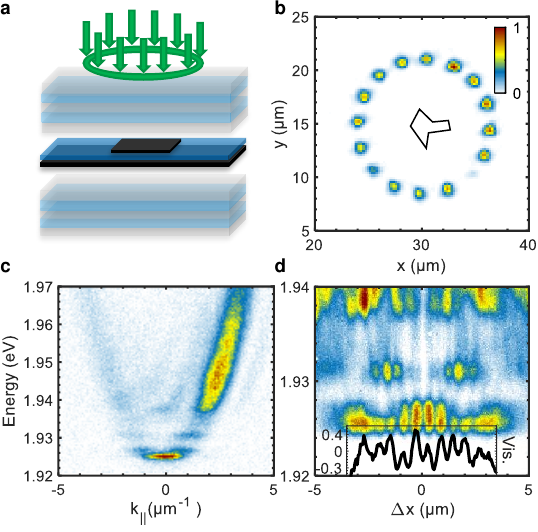}
\caption{(a) Schematics of the ring-shaped excitation around the small monolayer. (b) Spatially resolved intensity map of the excitation with the outline of the small monolayer. (c) Angle-resolved PL spectrum of the trap area. (d) Spectrally-resolved Michelson interference map of the trapped polaritons and (inset) the normalized interference fringes.}
\label{fig:Fig5}
\end{figure}

A distinct advantage of our trap design is the ability to inject free polaritons in a single monolayer region outside the trap, e.g. with a ring-shaped laser profile (see Fig.~\ref{fig:Fig5}a). We do this by using an optical mask \cite{Dall2014} with multiple holes arranged in a ring, and imaging it onto the sample (Fig. \ref{fig:Fig5}b). Because the propagation distance of the free, travelling polaritons is limited mainly by their lifetime \cite{Wurdack2021}, a fraction of them reach the trap and are confined. The angle-resolved PL, spatially filtered at the position of the trap, shows the free polariton emission above $E\approx1.94~\mathrm{eV}$, as well as the trapped polariton emission (see Fig.~\ref{fig:Fig5}c) with an inhomogeneous intensity distribution due to the cavity gradient. This demonstrates effective energy transfer through the Ga$_2$O$_3$ tunnel barrier by the polaritons that are exclusively excited on the large monolayer. The indirect excitation of trapped polaritons leads to the same magnitude ($g^{(1)}\approx 0.3$) and extent of spatial coherence (Fig.~\ref{fig:Fig5}d) as that observed when the pump is focused directly on the trap (see Fig.~\ref{fig:Fig4} and \cite{SM}), confirming the inherent partial coherence of the trapped polaritons in the thermal regime.

{\it Conclusion}.-- We have engineered a 2D trap for room-temperature WS$_2$ polaritons by locally increasing the light-matter coupling in a mechanically assembled, all-dielectric microcavity. We demonstrated that the confinement leads to enhancement of the ground state population due to more efficient energy redistribution of the trapped polariton population between the different energy states. Furthermore, we found significantly enhanced macroscopic phase coherence for the trapped polaritons compared to free polaritons, and linked it to suppression of inhomogeneous dephasing. Finally, our demonstration of populating the engineered trap indirectly, with free polaritons excited outside the trap, is an important step towards electrically-injected devices, e.g., polariton lasers utilising ring contacts \cite{Schneider2012}.




\begin{thebibliography}{9}
\bibitem{ANNF} The fabrication of the high-reflectivity substrate was performed at the ACT OptoFab node of the Australian National Fabrication Facility - a company established under the National Collaborative Research Infrastructure Strategy to provide nano and microfabrication facilities for Australia's researchers.
\end{thebibliography}
\end{document}